\documentclass[10pt,letterpaper]{article}

\usepackage[top=0.85in,left=2.75in,footskip=0.75in]{geometry}
\usepackage{amsmath,amssymb}
\usepackage{changepage}
\usepackage[utf8x]{inputenc}
\usepackage{textcomp,marvosym}
\usepackage{cite}
\usepackage{nameref,hyperref}
\usepackage[right]{lineno}
\usepackage{microtype}
\DisableLigatures[f]{encoding = *, family = * }
\usepackage[table]{xcolor}
\usepackage{array}

\newcolumntype{+}{!{\vrule width 2pt}}

\newlength\savedwidth

\usepackage{setspace} 
\raggedright
\usepackage[aboveskip=1pt,labelfont=bf,labelsep=period,justification=raggedright,singlelinecheck=off]{caption}

\makeatletter
\renewcommand{\@biblabel}[1]{\quad#1.}
\makeatother

\date{}

\usepackage{lastpage,fancyhdr,graphicx}
\usepackage{epstopdf}
\fancyheadoffset[L]{2.25in}
\fancyfootoffset[L]{2.25in}
\begin{document}
\vspace*{0.2in}

\begin{flushleft}
{\Large
\textbf\newline{A Systematic Identification and Analysis of Scientists on Twitter}
}
\newline
\\
Qing Ke\textsuperscript{1,*},
Yong-Yeol Ahn\textsuperscript{1},
Cassidy R. Sugimoto\textsuperscript{1},
\\
\bigskip
\textbf{1} School of Informatics and Computing, Indiana University, Bloomington, Indiana, US
\\
\bigskip
* qke@indiana.edu
\end{flushleft}

\section*{Abstract}
Metrics derived from Twitter and other social media---often referred to as
altmetrics---are increasingly used to estimate the broader social impacts of
scholarship. Such efforts, however, may produce highly misleading results, as
the entities that participate in conversations about science on these platforms
are largely unknown. For instance, if altmetric activities are generated mainly
by scientists, does it really capture broader social impacts of science? Here
we present a systematic approach to identifying and analyzing scientists on
Twitter. Our method can identify scientists across many disciplines,
without relying on external bibliographic data, and be easily adapted to
identify other stakeholder groups in science. We investigate the demographics,
sharing behaviors, and interconnectivity of the identified scientists. We find
that Twitter has been employed by scholars across the disciplinary spectrum,
with an over-representation of social and computer and information scientists;
under-representation of mathematical, physical, and life scientists; and a
better representation of women compared to scholarly publishing. Analysis of
the sharing of URLs reveals a distinct imprint of scholarly sites, yet only a
small fraction of shared URLs are science-related. We find an assortative
mixing with respect to disciplines in the networks between scientists,
suggesting the maintenance of disciplinary walls in social media. Our work
contributes to the literature both methodologically and conceptually---we
provide new methods for disambiguating and identifying particular actors on
social media and describing the behaviors of scientists, thus providing
foundational information for the construction and use of indicators on the
basis of social media metrics.

\section*{Introduction}
Twitter and other social media have become important communication channels for
the general public. It is thus not surprising that various stakeholder groups
in science also participate on these platforms. Scientists, for instance, use
Twitter for generating research ideas and disseminating and discussing
scientific results~\cite{Darling-role-2013, Faulkes-postpub-2014,
Woolston-flaw-2015}. Many biomedical practitioners use Twitter for engaging in
continuing education (e.g., journal clubs on Twitter) and other community-based
purposes~\cite{Lulic-physicians-2013}. Policy makers are active on Twitter,
opening lines of discourse between scientists and those making policy on
science~\cite{Kapp-policy-2015}.

Quantitative investigations of scholarly activities on social media---often
called altmetrics---can now be done at scale, given the availability of APIs on
several platforms, most notably Twitter~\cite{Priem-altwild-2012}. Much of the
extant literature has focused on the comparison between the amount of online
attention and traditional citations collected by publications, showing low
levels of correlation. Such low correlation has been used to argue that
altmetrics provide alternative measures of impact, particularly the broader
impact on the society~\cite{Bornmann-altoverview-2014}, given that social media
provide open platforms where people with diverse backgrounds can engage in
direct conversations without any barriers. However, this argument has not been
empirically grounded, impeding further understanding of the validity of
altmetrics and the broader impact of articles.

A crucial step towards empirical validation of the broader impact claim of
altmetrics is to identify scientists on Twitter, because altmetric activities
are often assumed to be generated by ``the public" rather than scientists,
although it is not necessarily the case. To verify this, we need to be able to
identify scientists and non-scientists. Although there have been some attempts,
they suffer from a narrow disciplinary focus~\cite{Haustein-astro-2014,
Hadgu-identify-2014, Holmberg-astrocomm-2014} and/or small
scale~\cite{Haustein-astro-2014, Holmberg-astrocomm-2014, Holmberg-diff-2014}.
Moreover, most studies use purposive sampling techniques, pre-selecting
candidate scientists based on their success in other sources (e.g., highly
cited in Web of Science), instead of organically finding scientists on the
Twitter platform itself. Such reliance on bibliographic databases binds these
studies to traditional citation indicators and thus introduces bias. For
instance, this approach overlooks early-career scientists and favors certain
disciplines.

Here we present the first large-scale and systematic study of scientists across
many disciplines on Twitter. As our method does not rely on external
bibliographic databases and is capable of identifying any user types that are
captured in Twitter list, it can be adapted to identify other types of
stakeholders, occupations, and entities. Our study serves as
a basic building block to study scholarly communication on Twitter and the
broader impact of altmetrics.

\section*{Background}
We classify current literature into two main categories, namely \emph{product}-
vs. \emph{producer-}centric perspectives. The former examines the sharing of
scholarly papers in social media and its impact, the latter focuses on who
generates the attention.

{\bf Product-centric perspective.} Priem and Costello formally defined Twitter
citations as ``direct or indirect links from a tweet to a peer-reviewed
scholarly article online" and distinguished between first- and second-order
citations based on whether there is an intermediate web page mentioning the
article~\cite{Priem-scholar-2010}. The accumulation of these links, they
argued, would provide a new type of metric, coined as ``altmetrics," which
could measure the broader impact beyond academia of diverse scholarly
products~\cite{Priem-altmetrics-2010}.

Many studies argued that only a small portion of research papers are mentioned
on Twitter~\cite{Priem-altwild-2012, Zahedi-altmetrics-2014,
Costas-altmetrics-2014, Hammarfelt-altmetrics-2014, Haustein-biotweet-2014,
Haustein-arxiv-2014, Haustein-character-2015}. For instance, a systematic study
covering $1.4$ million papers indexed by both PubMed and Web of Science found
that only $9.4\%$ of them have mentions on Twitter~\cite{Haustein-biotweet-2014},
yet this is much higher than other social media metrics except Mendeley.
The coverages vary across disciplines---medical and social sciences papers that
may be more likely to appeal to a wider public are more likely to be covered on
Twitter~\cite{Costas-thematic-2015, Haustein-character-2015}. Mixed results
have been reported regarding the correlation between altmetrics and
citations~\cite{Eysenbach-tweetation-2011, Shuai-arxiv-2012,
Thelwall-altmetrics-2013, Haustein-biotweet-2014, Winter-pone-2015}. A recent
meta-analysis showed that the correlation is negligible
($r=0.003$)~\cite{Bornmann-altmetrics-review-2015}; however, there is dramatic
differences across studies depending on disciplines, journals, and time window.

{\bf Producer-centric perspective.} Survey-based studies examined how scholars
present themselves on social media~\cite{Rowlands-media-2011,
Haustein-bib-2014, Loeb-urology-2014, Noorden-scientist-2014,
Bowman-diff-2015}. A large-scale survey with more than $3,500$ responses
conducted by \emph{Nature} in $2014$ revealed that more than $80\%$ were aware
of Twitter, yet only $13\%$ were regular users~\cite{Noorden-scientist-2014}.

A handful of studies analyzed how Twitter is used by scientists. Priem and
Costello examined $28$ scholars to study how and why they share scholarly
papers on Twitter~\cite{Priem-scholar-2010}. An analysis of $672$ emergency
physicians concluded that many users do not connect to their colleagues while a
small number of users are tightly interconnected~\cite{Lulic-physicians-2013}.
Holmberg and Thelwall selected researchers in $10$ disciplines and found clear
disciplinary differences in Twitter usages, such as more retweets by
biochemists and more sharing of links for economists~\cite{Holmberg-diff-2014}.

Note that these studies first selected scientists outside of Twitter and then
manually searched their Twitter profiles. Two limitations thus exist for these
studies. First, the sample size is small due to the nature of manual
searching~\cite{Priem-scholar-2010, Lulic-physicians-2013,
Chretien-physicians-2011, Haustein-astro-2014, Holmberg-diff-2014}. Second, the
samples are biased towards more well-known scientists. One notable exception is
a study by Hadgu and Jäschke, who presented a supervised learning based
approach to identifying researchers on Twitter, where the training set contains
users who were related to some computer science conference
handles~\cite{Hadgu-identify-2014, Pujari-researchers-2015}. Although this
study used a more systematic method, it still relied on the DBLP, an external
bibliographic dataset for computer science, and is confined to a single
discipline.

\section*{Identifying Scientists} \label{sec:identifying}
\subsection*{Scientist Occupations} \label{subsec:scientist-title}

Defining science and scientists is a Herculean task and beyond the scope of
this paper. We thus adopt a practical definition, turning to the $2010$
Standard Occupational Classification (SOC) system
(\url{http://www.bls.gov/soc/}) released by the Bureau of Labor Statistics,
United States Department of Labor. We use SOC because not only it is a
practical and authoritative guidance for the definition of scientists but also
many official statistics (e.g., total employment of social scientists) are
released according to this classification system. SOC is a hierarchical system
that classifies workers into $23$ major occupational groups, among which we are
interested in two, namely (1) Computer and Mathematical Occupations (code
15-0000) and (2) Life, Physical, and Social Science Occupations (code 19-0000).
Other groups, such as Management Occupations (code 11-0000) and Community and
Social Service Occupations (code 21-0000), are not related to science
occupations. From the two groups, we compile $28$ scientist occupations
(S1~Table). Although authoritative, the SOC does not always meet our intuitive
classifications of scientists. For instance, ``biologists" is not presented in
the classification. We therefore consider another source---Wikipedia---to
augment the set of scientist occupations. In particular, we add the occupations
listed at \url{http://en.wikipedia.org/wiki/Scientist\#By\_field}.

We then compile a list of scientist titles from the two sources. This is done
by combining titles from SOC, Wikipedia, and illustrative examples under each
SOC occupation. We also add two general titles: ``scientists" and
``researchers." For each title, we consider its singular form and the core
disciplinary term. For instance, for the title ``clinical psychologists," we
also consider ``clinical psychologist," ``psychologists," and ``psychologist."
We assemble a set of $322$ scientist titles using this method (S1~Data).

\subsection*{List-based Identification of Scientists}

\begin{figure}[!h]
\includegraphics[width=\textwidth]{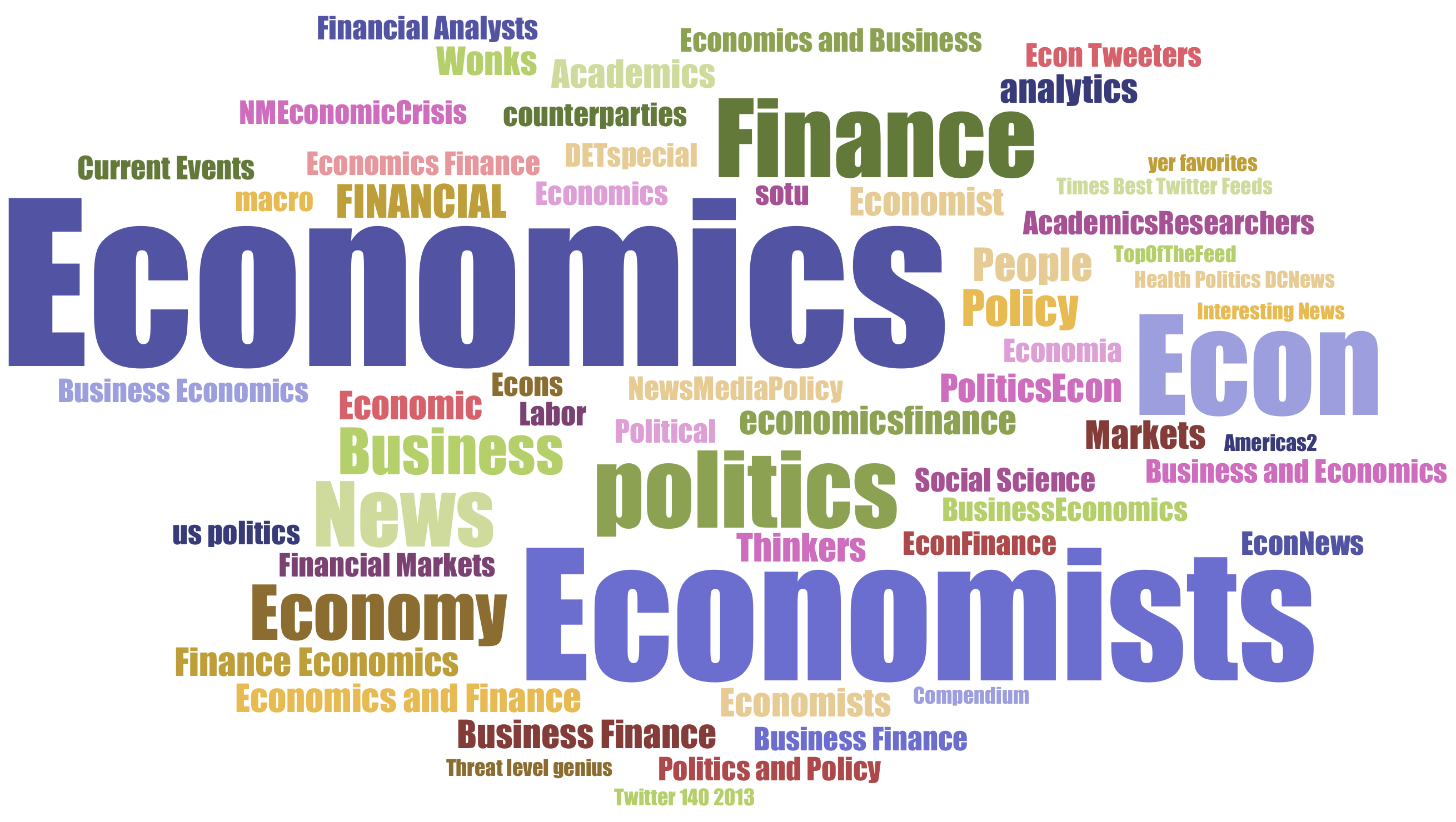}
\caption{{\bf User identity recored from Twitter list names.}
We show the word cloud of Twitter lists containing
\href{http://twitter.com/BetseyStevenson}{@BetseyStevenson}.}
\label{fig:list-name-wordcloud}
\end{figure}

Our method of identifying scientists is inspired by a previous study that used
Twitter \emph{lists} to identify user expertise~\cite{Sharma-whoiswho-12}. A
Twitter \emph{list} is a set of Twitter users that can be created by any
Twitter user. The creator of a list needs to provide a name and optional
description. Although the purpose of lists is to help users organize their
subscriptions, the names and descriptions of lists can be leveraged to infer
attributes of users in the lists. Imagine a user creating a list called
``economist" and putting
\href{http://twitter.com/BetseyStevenson}{@BetseyStevenson} in it; this signals
that @BetseyStevenson may be an economist. If @BetseyStevenson is included in
numerous lists all named ``economist," which means that many independent
Twitter users classify her as an economist, it is highly likely that
@BetseyStevenson is indeed an economist. This is illustrated in
Fig~\ref{fig:list-name-wordcloud} where the word cloud of the names of Twitter
lists containing @BetseyStevenson is shown. We can see that ``economist" is a
top word frequently appeared in the titles, signaling the occupation of this
user. In other words, we ``crowdsource" the identity of each Twitter user.

In principle, we could use Twitter's \texttt{memberships} API
(\url{https://dev.twitter.com/rest/reference/get/lists/memberships}), for each
user, to get all the lists containing this user, and then infer whether this
user is a scientist by analyzing the names and descriptions of these lists.
However, this method is highly infeasible, because (1) most users are not
scientists, (2) the distribution of listed counts is right-skewed: Lady~Gaga,
for example, is listed more than $237K$ times
(\url{https://www.electoralhq.com/twitter-users/most-listed}), and (3) Twitter
API has rate limits. We instead employ a previously introduced list-based
snowball sampling method~\cite{Wu-saywhat-2011} that starts from a given
initial set of users and expands to discover more. We improve this approach by
more systematically obtaining the job title lexicon, as described in the last
section. Moreover, instead of choosing a few preselected users, we obtain a
total of $8,545$ seed users by leveraging the results of a previous work that
identified user attributes using Twitter lists~\cite{Sharma-whoiswho-12}
(S1~Text).

We use the snowball sampling (breadth-first search) on Twitter lists. We first
identify seed users (S1~Text) and put them into a queue. For each public user
in the queue, we get all the lists in which the user appears, using the Twitter
\texttt{memberships} API. Then, for each public list in the subset resulting
lists whose name contains at least one scientist title, we get its members
using the Twitter \texttt{members} API
(\url{https://dev.twitter.com/rest/reference/get/lists/members}) and put those
who have not been visited into the queue. The two steps are repeated until the
queue is empty, which completes the sampling process. Note that to remove many
organizations and anonymous users as well as to speed up the sampling, we only
consider users whose names contain spaces. We acknowledge that this may drop
many users with non-English names or the ones who do not disclose their names
in a standard way. Also note that this procedure is inherently blind towards
those scientists who are not listed.

From the sampling procedure, we get $110,708$ users appearing in $4,920$ lists
whose names contain scientist titles. To increase the precision of our method,
the final dataset contains those users whose profile descriptions also contain
scientist titles. A total number of $45,867$ users are found.

\section*{Analyzing Scientists} \label{sec:analysis}
For each of the $45,867$ identified scientists, we obtain their followers,
followings, and up to $3,200$ most recent statuses (tweets, retweets, and
replies) using Twitter APIs. In total, we get $88,412,467$ following pairs and
$64,449,234$ statuses. With this dataset, we ask the following questions:
\begin{itemize}
\item What are the demographics of identified scientists on Twitter, in terms
of discipline and gender?
\item What are the URLs scientists post in their tweets?
\item How do scientists follow/retweet/mention each other on Twitter and who
are the most ``influential" scientists in these interactions?
\end{itemize}
These questions are necessary for the validation and appropriate utilization of
altmetrics for research evaluation.

\subsection*{Who are they?} \label{subsec:analysis-demo}

We investigate the demographics of identified scientists in terms of discipline
and gender.

\paragraph*{Discipline.} In contrast to previous analyses that either focused
on a single discipline~\cite{Haustein-astro-2014, Hadgu-identify-2014,
Holmberg-astrocomm-2014} and/or relied on a small number of accounts in a few
disciplines~\cite{Haustein-astro-2014, Holmberg-astrocomm-2014,
Holmberg-diff-2014}, our systematic approach covers a wide range of
disciplines, thus allowing us to investigate the representativeness of
scientists in different disciplines. Moreover, identifying disciplines also
allows us to analyze behavioral differences by disciplines and understand
inter-disciplinary interactions between scientists.

To identify the discipline of each scientist, we leverage the compiled list of
scientist titles. They are searched in profile descriptions and in assigned
list names. Whereas profiles provide us information about how scientists
perceive themselves, list names tell us how they are perceived by others. When
searching, we begin with longer names and then move on to shorter ones. For
instance, ``I am an evolutionary biologist" will be matched with ``evolutionary
biologist" not with ``biologist." When multiple matches are found, each of them
will be counted once. From profile descriptions, we obtain a total of $25,798$
($56.2\%$) users whose profile descriptions contain at least one scientist
title, suggesting that a majority of ``perceived" scientists identify
themselves as scientists. S2~Table shows the number of users for each of the
top $30$ scientist titles extracted from profile descriptions. Psychologists
are the most numerous, which may be rooted in two reasons. First, many types of
psychology practitioners (e.g., counseling psychologists) are presented in the
scientist titles. Second, many of them may not be resident in academia and
serve as health care professionals. Thus, they may show this in their profiles
to signal their profession. Clinical psychologists, for instance, are also
highly represented. Other common type of scientists include physicists,
computer scientists, and archaeologists.

When extracting titles from list names, there are $24,635$ ($53.1\%$) users who
are included in at least one list whose name contain scientist titles. S3~Table
presents the number of users for each title. We observe some differences
between the two rankings. Computer scientists, for instance, fail to make it
into the top $10$ based on list names, indicating that they are less often to
be labeled by other users as ``computer scientists" instead as other labels
(e.g., ``data scientists"). Sociologists, on the other hand, show the opposite
trend.

Based on the titles extracted from profiles and list names, we now assign each
user a final title or titles. We give more weight to titles from profiles by
using profile information first when they are available. If this fails, we
choose the title that appears the most times in the lists. With this procedure,
we assign disciplines to $30,793$ ($67.1\%$) users.
Table~\ref{tab:top-disciplines} shows the number of users in the top $24$
disciplines. These results again demonstrate that our method can discover
scientists from diverse disciplines of sciences and social sciences.

\begin{table}[!ht]
\centering
\caption{{\bf Number of users in most presented disciplines.}}
\begin{tabular}{l r | l r}
\hline
Discipline          & Users & Discipline             & Users \\ \hline
Historian           & 3586  & Ecologist              & 775 \\
Psychologist        & 3579  & Anthropologist         & 698 \\
Physicist           & 2737  & Astronomer             & 675 \\
Nutritionist        & 2510  & Statistician           & 619 \\
Political scientist & 1441  & Clinical psychologist  & 576 \\
Computer scientist  & 1123  & Linguist               & 526 \\
Archaeologist       & 1100  & Social scientist       & 438 \\
Biologist           & 1075  & Geographer             & 430 \\
Economist           & 1044  & Epidemiologist         & 403 \\
Sociologist         & 1020  & Mathematician          & 370 \\
Neuroscientist      & 916   & Geologist              & 359 \\
Meteorologist       & 855   & Evolutionary biologist & 330 \\
\hline
\end{tabular}
\label{tab:top-disciplines}
\end{table}

We investigate whether some disciplines are over- or under-represented in
social media by comparing the results in Table~\ref{tab:top-disciplines} with
the size of the science workforce. To do so, we use the total employment data
from the latest (May 2014) National Occupational Employment Statistics (OES;
\url{http://www.bls.gov/oes/current/oes_nat.htm}), which lists the size of
workforce for each occupation. We aggregate the number of scientists onto the
OES minor level (computer and information, mathematical, life, physical, and
social scientists), and Table~\ref{tab:discipline-comparison} shows the total
number and the percentage of employment for each OES minor group as well as
results from Twitter. These results suggest that social scientists and computer
and information scientists are over-represented on Twitter, whereas
mathematical, life, and physical scientists are under-represented. We should,
however, note that (1) this is a rough estimation, as OES is solely for US but
users in our sample may come from other countries, and (2) the results could
also be biased due to our list-based sampling method. Therefore, further work
is needed to check whether our results reflect an accurate representation on
Twitter.

\begin{table}[!ht]
\centering
\caption{{\bf Comparing number of scientists on Twitter and the size of the
science workforce.}}
\begin{tabular}{l r r r r}
\hline
Title             & Employment & Employment $\%$ & Twitter $\%$ & Ratio   \\ \hline
Computer \& Info. & $24,210$   & $2.71\%$        & $3.62\%$     & $1.336$ \\
Mathematical      & $138,540$  & $15.48\%$       & $3.18\%$     & $0.205$ \\
Life              & $269,660$  & $30.13\%$       & $25.18\%$    & $0.836$ \\
Physical          & $274,520$  & $30.68\%$       & $19.66\%$    & $0.641$ \\
Social            & $187,910$  & $21.00\%$       & $48.37\%$    & $2.303$ \\
\hline
\end{tabular}
\label{tab:discipline-comparison}
\end{table}

\paragraph*{Gender.} To identify gender, we first remove two common prefixes
(``Dr." and ``Prof.") and then search the first names in the $1990$ US census
database of frequently occurring first names
(\url{http://www.census.gov/topics/population/genealogy/data/1990_census/1990_census_namefiles.html}),
resulting in $11,910$ females and $18,882$ males. For the remaining unknown
users, we detect their gender by using a facial feature detection service
provided by Face++ (\url{http://www.faceplusplus.com/detection_detect/}). The
input of this service is the URL of the image, which we use the profile image
URL provided by Twitter, and one of its returned values is the gender
associated with a confidence value. We only keep the gender results with
confidence greater than $90$. Combining the two methods, we are able to
identify the gender for $71.9\%$ ($12,732$ females and $20,232$ males) of the
sample. Of those identified, $38.6\%$ were female and $61.4\%$ were male, or
the female to male ratio is $0.629$.

We compare the female-male ratio of the sampled scientists on Twitter to the
ones derived from two other samples, namely general Internet users and offline
scientific authorships. A recent report from the Pew Research Center shows that
$21\%$ and $24\%$ of female and male Internet users use Twitter
(\url{http://www.pewinternet.org/2015/01/09/demographics-of-key-social-networking-platforms-2/}),
leading to the female-male ratio $21\% / 24\% = 0.875$ (the ratio for Internet
users is 0.99.). Regarding scientific authorships, the ratio ranges from
$0.179$ (Iran) to $0.754$ (Poland), and is $0.428$ for
US~\cite{Lariviere-gender-2013}. Based on these, the gender ratio is less
skewed for scientists on Twitter compared with scientific authorships in US,
supporting the argument that Twitter provides more opportunities for diverse
participation from women.

\subsection*{What do they share?} \label{subsec:analysis-share}

\begin{figure}[!h]
\includegraphics[width=\textwidth]{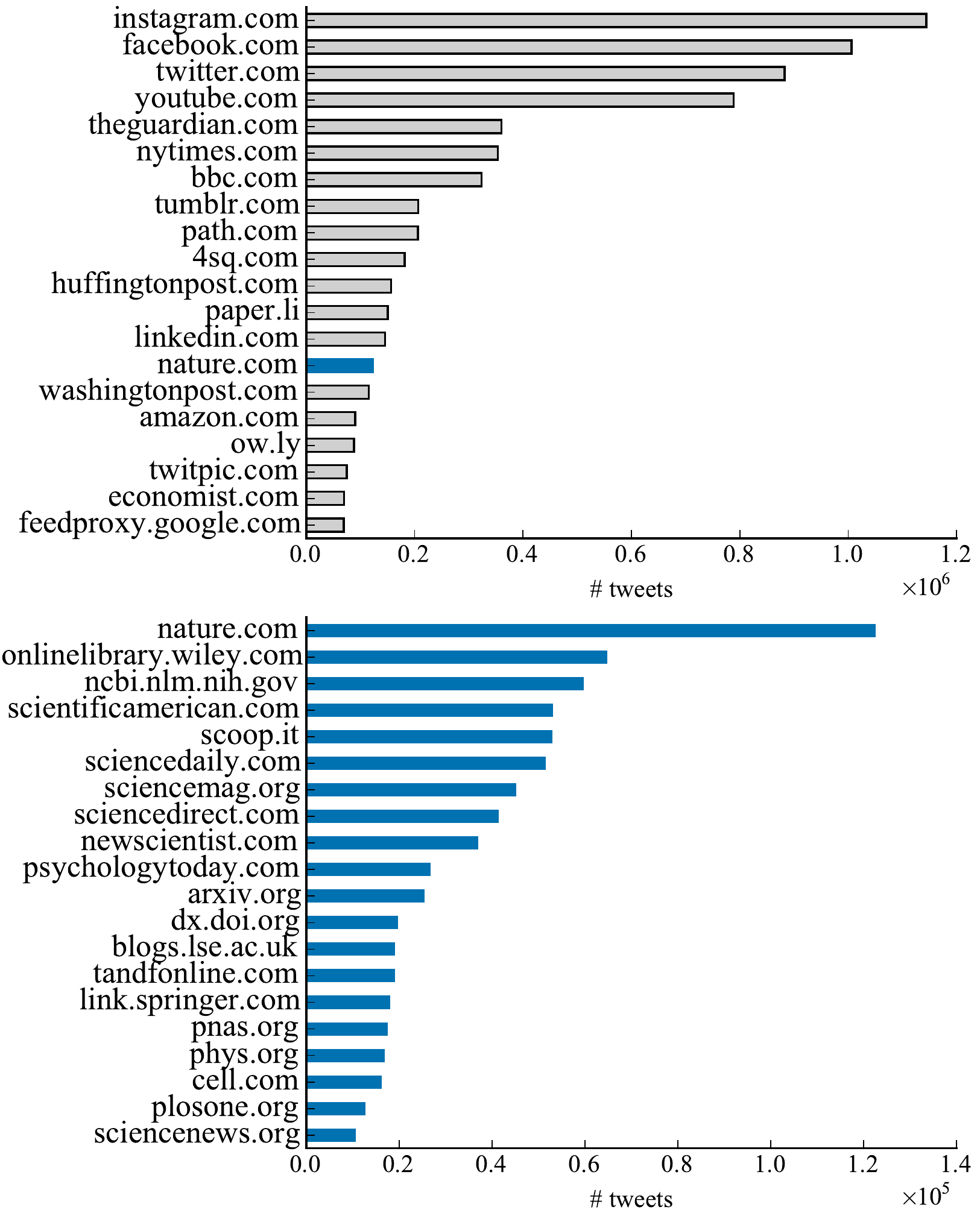}
\caption{{\bf Top 20 domains.}
We extract URLs from tweets and retweets and then count the appearances of the
domains. Top: overall. Bottom: scientific.}
\label{fig:top-domains}
\end{figure}

We study tweet contents posted by scientists. We specifically focus on URLs to
understand sharing of scientific articles on Twitter. To do so, we extract URLs
from tweets and retweets, ignoring replies. We only consider those generated
from the retweet button as retweets and extract URLs from their original
tweets. Noting that many top domains are shortened URLs (e.g., bit.ly), we
expand them and extract domain names. Fig~\ref{fig:top-domains} (top) shows the
top $20$ domains and number of tweets mentioning them. We observe that many of
them are to news websites, such as \emph{The Guardian} and \emph{The New York
Times} and the domain for the Nature Publishing Group also ranks in
the top. Fig~\ref{fig:top-domains} (bottom)
displays the top scientific domains. Major academic publishers, such as Wiley
(onlinelibrary.wiley.com), Elsevier (sciencedirect.com), Taylor $\&$ Francis
(tandfonline.com), and Springer (link.springer.com) appear in the top. Journals
like \emph{Science}, \emph{PNAS}, and \emph{PLoS ONE} also attract much
attention on Twitter.

\begin{figure}[!h]
\includegraphics[width=\textwidth]{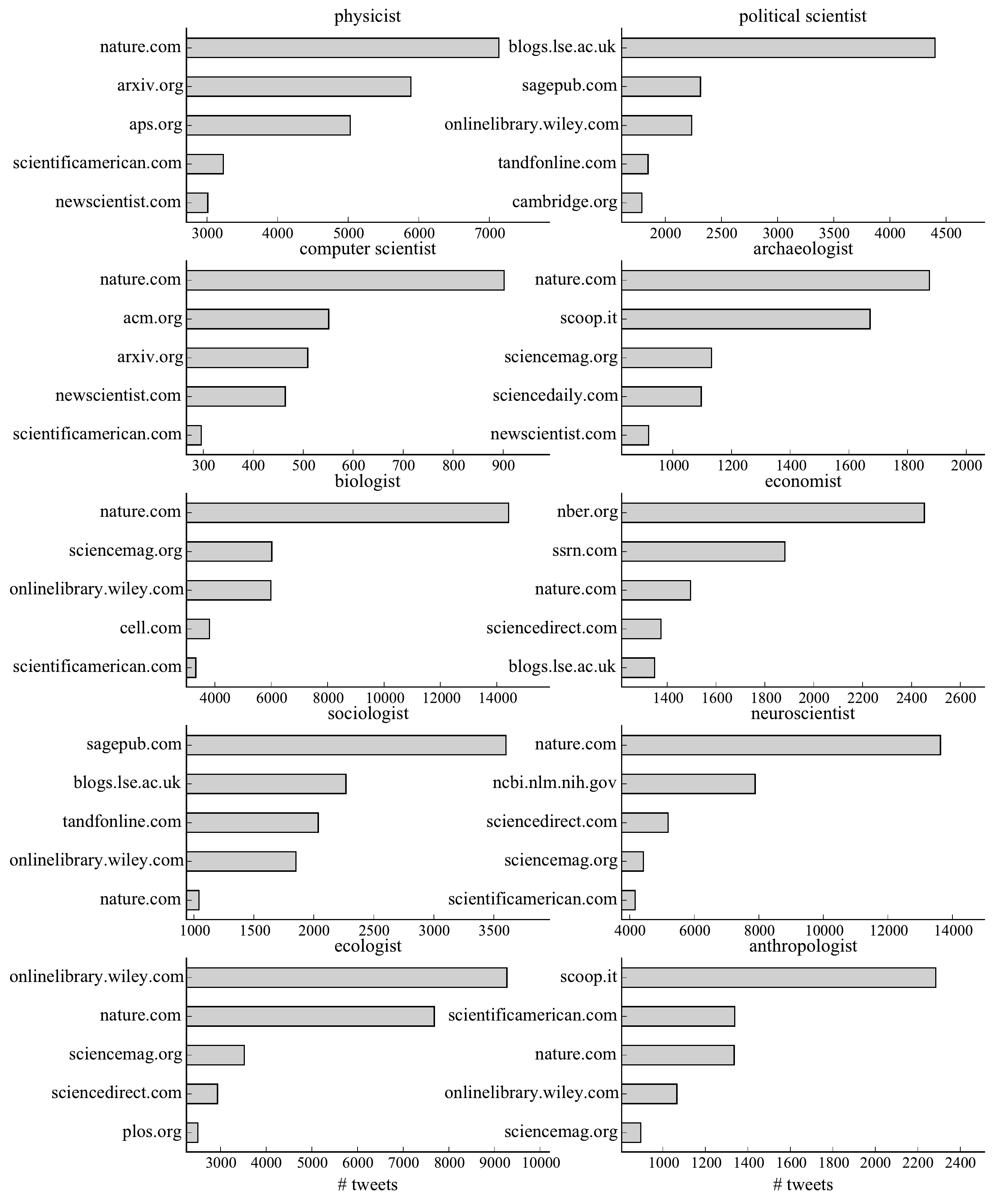}
\caption{{\bf Top scientific domains by disciplines.}
We extract URLs from tweets and retweets and then count the appearances of the
scientific domains in each of the $10$ disciplines.}
\label{fig:top-domains-discipline}
\end{figure}

To understand disciplinary differences of the posted URLs,
Fig~\ref{fig:top-domains-discipline} shows the top $5$ scientific domains
shared by scientists in each discipline. Although some domains such as
\url{nature.com} are popular across disciplines, scientists are more likely to
share content from their disciplines. For instance, \url{arxiv.org}, a
pre-print server mainly for physics, and \url{aps.org}, the website for the
American Physical Society, are the top domains for physicists. \url{acm.org} is
popular among computer scientists. The blog for the London School of Economics
and Political Science (LSE) (\url{http://blogs.lse.ac.uk}) is popular among
political scientists, economists, and sociologists.

\begin{figure}[!h]
\includegraphics[width=\textwidth]{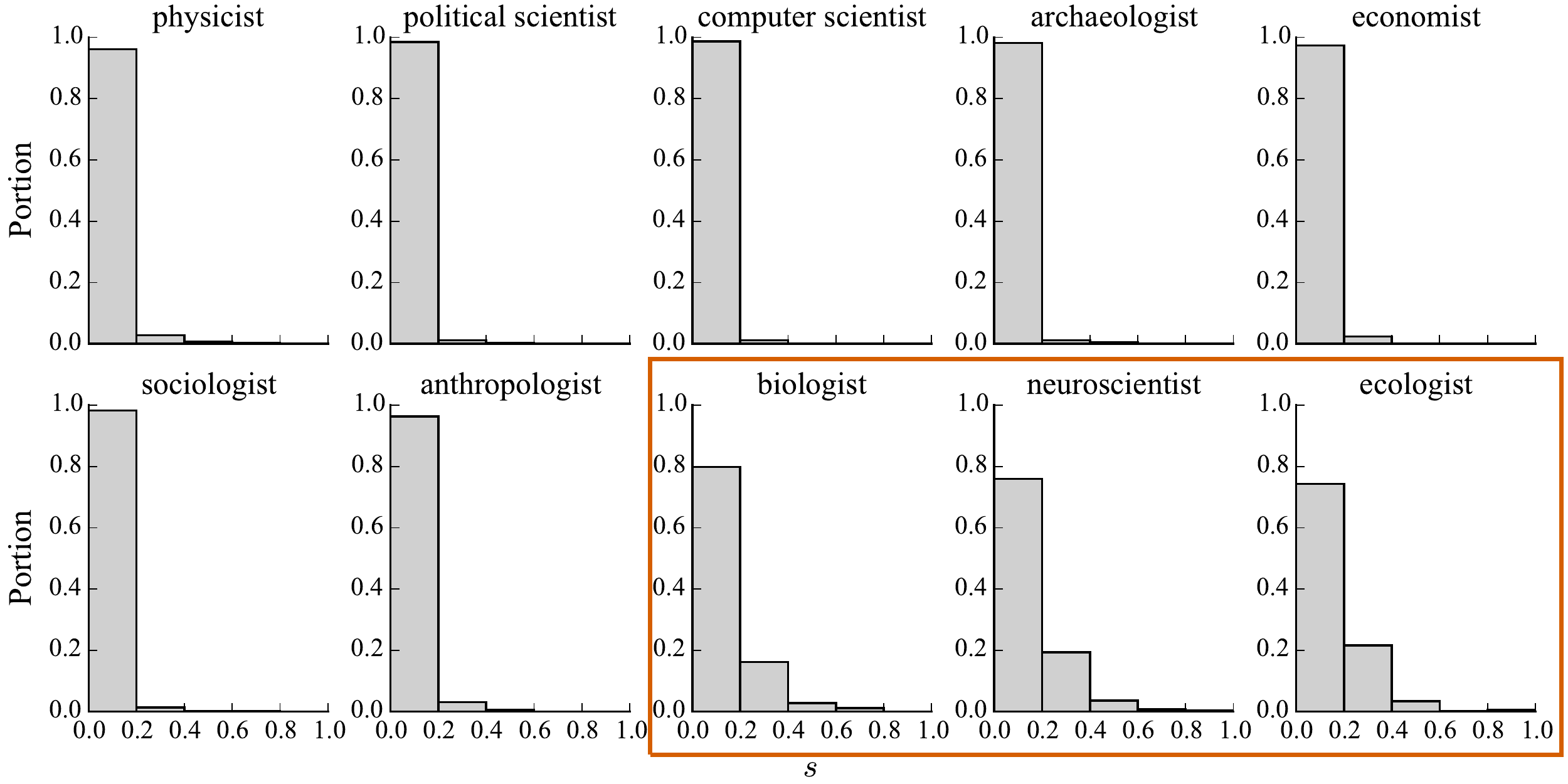}
\caption{{\bf Distribution of fraction of scientific domains.}
For each scientist, $s$ is the fraction of (re)tweets that contains URLs
referring to scientific websites to the total number of (re)tweets that
contains URLs. We show the histograms of $s$ of scientists for each of the $10$
disciplines.}
\label{fig:scientific-domain-fraction}
\end{figure}

To understand to what extent scientists share scientific URLs, we calculate for
each user the fraction $s$ of (re)tweets that contains URLs referring to
scientific websites to the total number of (re)tweets that contains URLs.
Fig~\ref{fig:scientific-domain-fraction} shows histograms of $s$ by
disciplines. Clearly, the fractions are small across all disciplines, while
biological scientists---biologists, neuroscientists, and ecologists---post more
tweets referring to scientific domains. For other types of scientists, the
fraction is smaller than $0.2$ for nearly all of them. This suggests that for
most scientists on Twitter, sharing links to scientific domains is a minor
activity.

\subsection*{How do they connect to each other?} \label{subsec:analysis-connectivity}

\begin{figure}[!h]
\includegraphics[trim=8mm 5mm 2mm 10mm, clip, width=\textwidth]{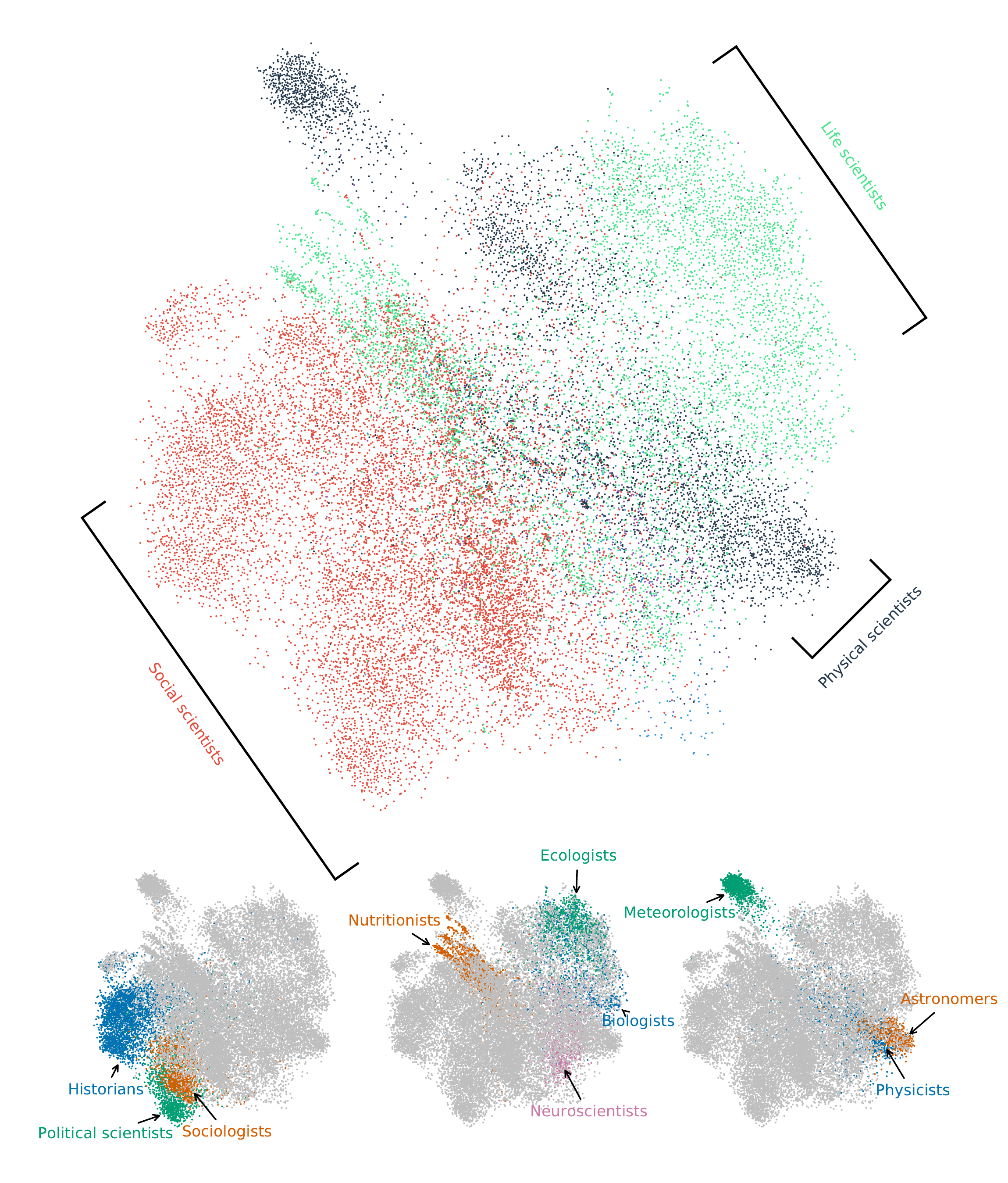}
\caption{{\bf Follower network of scientists on Twitter.}
We use the ForceAtlas2~\cite{Jacomy-ForceAtlas2-2014} algorithm to layout the
largest connected component of the follower network with mutual following
relations. We only show nodes with known disciplines.}
\label{fig:follower-net}
\end{figure}

We investigate how scientists connect with each other, by examining the
follower, retweet, and mention networks between them. In the follower network,
a directed and unweighted link from user $a$ to $b$ means that $a$ follows $b$.
In the retweet network, a directed link pointing from $a$ to $b$ is weighted,
with the weight representing the number of times that $a$ has retweeted $b$'s
tweets. In the mention network, a link is also directed and weighted, and the
weight indicates the number of times that $a$ has mentioned $b$ in $a$'s
tweets. Table~\ref{tab:network-stats} reports summary statistics of the largest
weakly connected components in the three networks. Fig~\ref{fig:follower-net}
shows the follower network, where each node is a scientist and the color
represents the extracted title.

\begin{table}[!ht]
\centering
\caption{{\bf Summary statistics of scientist networks.}}
\begin{tabular}{c l r r}
\hline
Network  & Links             & \# nodes & \# links \\ \hline
Follower & Who-follows-whom  & $39,485$ & $1,234,905$ \\
Retweet  & Who-retweets-whom & $30,204$ & $480,479$ \\
Mention  & Who-mentions-whom & $26,078$ & $168,232$ \\
\hline
\end{tabular}
\label{tab:network-stats}
\end{table}

\paragraph*{Centralities.} Given the three networks, we investigate the most
influential---operationalized by network centralities---scientists on Twitter.
S4~Table lists top scientists under the centrality of in-degree
$d_{\leftarrow}$ or in-strength $s_{\leftarrow}$, PageRank $PR$, and $k$-core
number $k$ in the three networks. We observe that there are some overlaps
between $d_{\leftarrow}$ ($s_{\leftarrow}$) and $PR$ across the three networks
and that top scientists in terms of $k$-core are different from those in terms
of the other two measures. Regarding individual users, Neil deGrasse Tyson
(\href{http://twitter.com/neiltyson}{@neiltyson}), an astrophysicist, is ranked
the first under degree and PageRank for all the three networks.

\begin{figure}[!h]
\includegraphics[width=\textwidth]{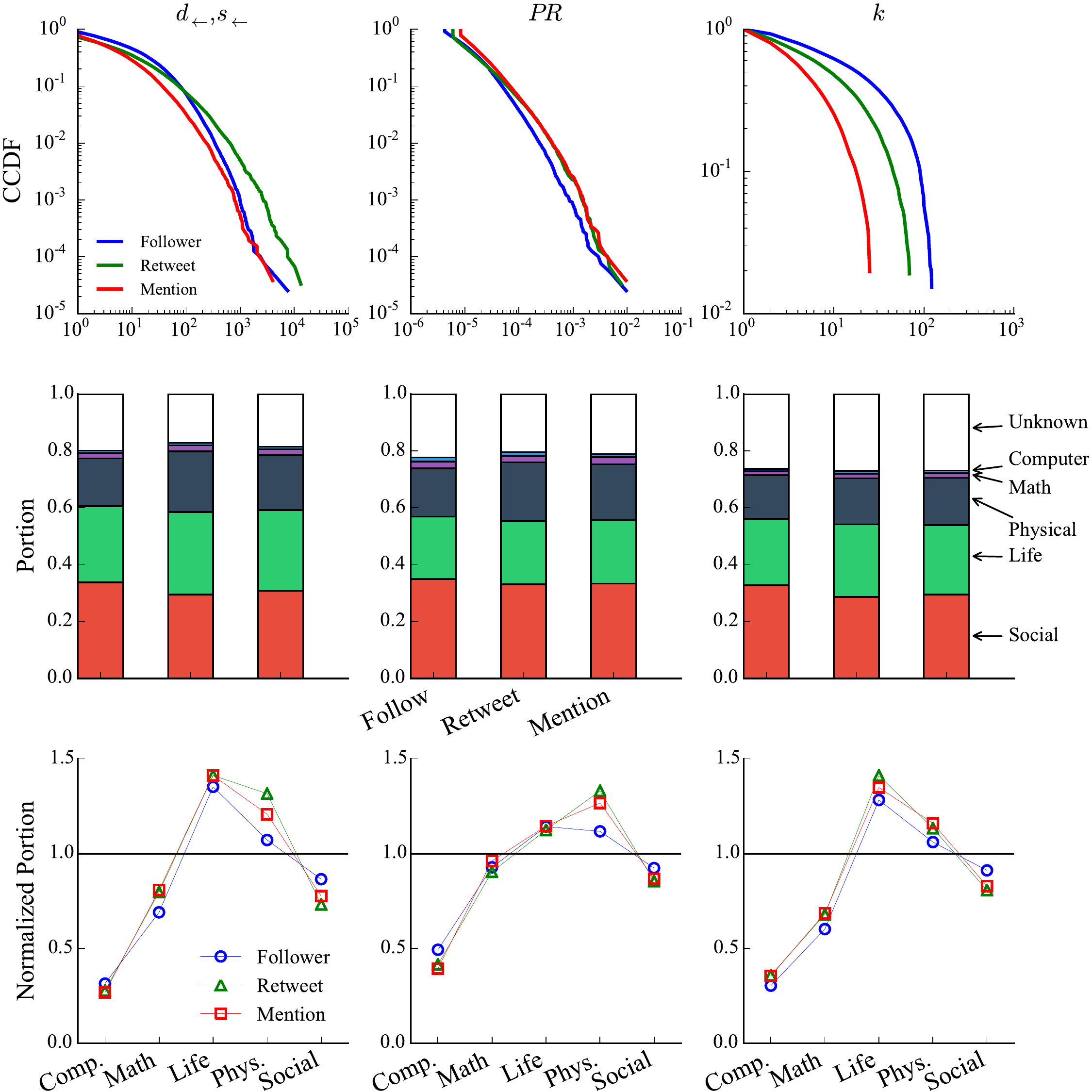}
\caption{{\bf Distributions of centralities in the follower, retweet, and
mention networks between scientists.}
Top: Distribution of in-degree $d_{\leftarrow}$ or in-strength
$s_{\leftarrow}$, PageRank ($PR$), and $k$-core number $k$ in the three
networks. Middle: Portion of centralities occupied by scientists in each group.
We calculate, for each centrality in each network, the sum of centrality values
of users in each scientist group divided by the total centrality values.
Bottom: Normalized portion of the three types of centralities occupied by
scientists in each group in the three networks. Normalization is done by
dividing the portion by number of scientists.}
\label{fig:net-centrality}
\end{figure}

Going beyond top nodes, we show in Fig~\ref{fig:net-centrality} (top) the
distributions of centralities in the three networks. We observe that the
distribution for $k$-core number is less heterogeneous than the other two
centralities across the three networks, and the distributions of PageRank are
similar for the three networks. The heterogeneity in centrality distributions
raises the question of how attention is distributed among disciplines. We thus
calculate, for each centrality in each network, the sum of centrality values of
users in each OES minor group (mathematical, life, etc.) divided by the total
values of the centrality. Fig~\ref{fig:net-centrality} (middle) shows results.
We can see that social and life scientists account for the largest part of
centralities, followed by physical scientists. Mathematical and computer
scientists only occupy a very small portion. Combining these results and
Table~\ref{tab:discipline-comparison}, we further ask how this can be explained
by the number of scientists in each minor group and whether scientists from
some groups disproportionately account for the centralities. For each
centrality in each network, we normalize the fraction of the centrality by the
fraction of scientist in each group. Fig~\ref{fig:net-centrality} (bottom)
shows the normalized portion of centralities possessed by each OES minor group,
where the result greater than one means that scientists in the group
disproportionately have larger centralities, which is the case for life and
physical scientists under all the three centralities in the three networks.
Computer, mathematical, and social scientists exhibit the opposite pattern.

\paragraph*{Assortativity.} Assortativity quantifies the tendency that nodes
with similar attributes are connected~\cite{Newman-assort-2003}. We observe
from Fig~\ref{fig:follower-net} that scientists from the same discipline are
positioned closely, signaling a positive assortativity with respect to
discipline. This is indeed the case, as the assortativity coefficient is
$0.548$ (ignoring nodes with unknown discipline), indicating that the follower
network is assortative by discipline---scientists tend to follow others from
their own discipline group. However, this is not the case for gender
(assortativity coefficient $0.054$), implying that scientists follow others
with the same gender not more often than expectation by pure chance. The
retweet and mention networks are also assortative with respect to disciplines,
with coefficient $0.492$ and $0.537$, but not to gender (coefficient $0.074$
and $0.086$).

\section*{Discussion}
Our work presents an improvement over earlier methods of identifying scientists
on Twitter by selecting a wider array of disciplines and extending the sampling
method beyond the paper-centric approach. Our method may serve as a useful step
towards more extensive and sophisticated analyses of scientists on Twitter---it
cannot be assumed that the population of scientists on Twitter is similar in
composition and behavior to the population of scientists represented in
traditional bibliometric databases. Therefore, sampling should be independent
of these external data and metrics. Furthermore, in seeding with terms from the
Standard Occupational Classification provided by the Bureau of Labor
Statistics, we are able to classify both scholarly and practitioner scientific
groups, thus widening the conceptualization of scientists on Twitter.

The triangulation of list- and bio-based classifications of scholars allows us
to integrate two perspectives on identity: how scientists self-identified and
how they were identified by the community. Our approach favors precision over
recall; that is, we feel confident that those identified were scientists, but
there is a much larger population of scientists who were not identified in this
way.

Our disciplinary analyses suggest that Twitter is employed by scholars across
the disciplinary spectrum---historians were widely represented, as were
physicists, political scientists, computer scientists, biologists, economists,
and sociologists. Practitioners were also highly represented---psychologists
and nutritionists were in the top five in terms of disciplines with the highest
number of identified members. However, a large percentage was also explicitly
academic scholars: self-identified students and faculty members comprised
$21.9\%$ of the total population (S1~Text). Our analysis suggests that social
scientists are overrepresented on Twitter, given their proportional
representation in the scientific workforce, and that mathematicians are
particularly underrepresented. Our findings resonate with some previous
results~\cite{Haustein-character-2015}, which looked at social media metric
coverage of publications by field. They found higher Twitter density in the
social and life sciences and lower density for mathematics and computer
science. This provides some intuitive alignment: if a group is systematically
underrepresented on the platform, we might expect a lower degree of activity
around papers within that discipline.

Of those whose gender could be identified, $38.6\%$ were female and $61.4\%$
were male. This represents a more equal representation of women than seen in
other statistics on the scientific workforce, such as number of
publications~\cite{Lariviere-gender-2013}, suggesting that Twitter scientists
may be more gender-balanced than the population of publishing scientists.

As might be expected, scientists tweet in much the same way as the general
population: Instagram, Facebook, YouTube are among the most tweeted domains,
along with general news sites such as \emph{The Guardian}, \emph{New York
Times}, and the \emph{BBC}. However, scientists also have a distinct imprint of
scholarly sites, such as generalists publications (i.e., \emph{Nature} and
\emph{Science}) and reinforce the academic oligarchy of journal
publishers~\cite{Lariviere-publishers-2015}. The popular pre-print server,
arXiv, also occupies a prominent spot among the top $20$ cited domains.
However, overall, tweets to these URLs identified as scientific only
represented a small fraction of the overall tweets, suggesting that the content
of scientists' tweets is highly heterogeneous. This reinforces previous
studies, which showed a strong blurring of boundaries between the personal and
professional on Twitter, under a single Twitter handle~\cite{Bowman-diff-2015}.

We operationalized centralities in three ways: by followers, retweets, and
mentions. Social and life scientists dominate these networks and mathematicians
and computer scientists are relatively isolated. However, once these
centralities are normalized by the size of the group, social scientists
actually underperform, given their size. This is imperative information for the
construction of indicators on the basis of these metrics. Just as it is
standard bibliometric practice to normalize by field, so too should altmetric
practices integrate normalization, given the uneven distribution of disciplines
represented on these platforms.

Analysis of assortativity suggests that disciplinary communities prevail in the
unfiltered realm of social media---scholars from the same disciplines tended to
follow each other. This could suggest a negative result in terms of broader
impact of social media metrics---if disciplinary walls are maintained in this
space, it may not provide the unfettered access to scholarship that was
promised. Furthermore, networks of communities reveal some isolation: e.g.,
although they represent a large proportion of the total users identified,
historians are largely isolated in the Twitter network.

Our work has the following limitations. First, the reliance of Twitter lists
leads to our method inherently blind towards those scientists who are not
listed. Furthermore, the use of lists may skew
towards the elite and high profile science communicators (e.g.,
Neil~deGrasse~Tyson). Second, in the sampling process, the exclusion of users
whose names are without spaces biases the sample towards English-speaking users
and causes many scientists not discovered. Third, the existence of private
lists prohibits us to get the members there and affects further discovery of
new users. Fourth, how list members were curated is largely unknown, and this
might be done automatically and thus decrease the precision of identified
scientists. Fifth, in the post-processing, the filtering of users whose profile
descriptions do not contain scientist titles biases the sample towards
self-disclosed scientists.

\section*{Conclusion}
In this work, we have developed a systematic method to discovering scientists
who are recognized as scientists by other Twitter users through Twitter list
and self-identify as scientists through their profile. We have studied the
demographics of identified scientists in terms of discipline and gender,
finding over-representation of social scientists, under-representation of
mathematical and physical scientists, and a better representation of women
compared to the statistics from scholarly publishing. We have analyzed the
sharing behaviors of scientists, reporting that only a small portion of shared
URLs are science-related. Finally, we find an assortative mixing with respect
to disciplines in the follower, retweet, and mention networks between
scientists.

Future work is needed to examine the use of machine learning
methods~\cite{Hadgu-identify-2014} by leveraging information from retweet and
mention networks to improve our identification method, to investigate the
degree to which a more equal representation of women is due to age, status, or
the representation of practitioners in our dataset, and to ascertain to what
extent altmetric communities (i.e., follow, retweet, and mention networks)
align with or differ from bibliometrically-derived communities (i.e., citation
and collaboration networks).

\section*{Supporting Information}

\paragraph*{S1 Text.}

\paragraph*{S1 Fig. Network of communities.}

\paragraph*{S1 Table. Scientist occupations from 2010 Standard Occupational
Classification released by US Department of Labor.}

\paragraph*{S2 Table. Top scientist titles from profile descriptions.}

\paragraph*{S3 Table. Top scientist titles from Twitter list names.}

\paragraph*{S4 Table. Top scientists in the follower, retweet, and mention
networks between scientists by in-degree $d_{\leftarrow}$ or in-strength
$s_{\leftarrow}$, PageRank ($PR$), and $k$-core number.}

\paragraph*{S5 Table. Top users in each community.}

\paragraph*{S1 Data. List of scientist titles.}

\section*{Acknowledgments}

We thank Onur Varol for early discussions and Filippo Radicchi for providing
the computing resource. YYA acknowledges support from Microsoft Research. CRS
is supported by the Alfred P. Sloan Foundation Grant \#G-2014-3-25.

\clearpage

\setcounter{figure}{0}
\makeatletter 
\renewcommand{\thefigure}{S\@arabic\c@figure}
\makeatother

\setcounter{table}{0}
\makeatletter 
\renewcommand{\thetable}{S\@arabic\c@table}
\makeatother

\newcommand{\sname}[1]{\href{http://twitter.com/#1}{#1}}

\begin{center}\LARGE\textbf{S1 Text}\end{center}

\textbf{Getting initial seed users:} One easy way to obtain the initial
seed users is to use an established set of scientists, for instance, the top
$100$ science stars
(\url{http://news.sciencemag.org/scientific-community/2014/10/twitters-science-stars-sequel})
compiled by \emph{Science}. However, this may introduce bias towards more
popular scientists and disciplines. Given our goal of identifying scientists at
the scale of the entire Twitter platform, we instead take a more systematic
approach by leveraging the results of a previous work that identified
attributes of Twitter users~\cite{Sharma-whoiswho-12}. The attributes of a user
are the most frequently used words in the names and descriptions of the lists
containing the user. These attributes are provided via the website
\url{http://twitter-app.mpi-sws.org/who-is-who} that takes the screen name of a
Twitter user as input and returns a word cloud for the given user with font
sizes of words encoding the frequency of their appearance in list names and
descriptions. Note that attributes are only available for those users who are
included in at least $10$ lists~\cite{Sharma-whoiswho-12}.

We first collect $285,760,507$ unique users by scanning a Twitter Gardenhose
dataset, which contains about $10\%$ of all public tweets from January $2013$
to June $2014$. The number of users is comparable to the number reported in a
previous large-scale Twitter study~\cite{Gabielkov-twitter-14}, and the set of
users covers any account that tweeted at least once and at least one of these
tweets is included in Gardenhose during the period. We then filter out those
users who were listed less than $8$ times in our corpus, and query all the
remaining users to the who-is-who website, finally obtaining attributes of
$2,436,889$ users.

We then obtain seed users who are most likely to be scientists from the $2.4M$
users. As the seeds will be used for expansion, we prefer precision to recall.
We thus adopt stringent criteria to filter out non-seed users. Specifically, we
disregard the least important attributes of each user and then keep those users
whose attributes contain the attribute ``science" and at least one scientist
title compiled before. The obtained initial set has $8,545$ users, and we use
them as initial seeds for snowball sampling.

\textbf{Academic rank:} It is also interesting to investigate academics and
to understand how scientists with different academic ranks (PhD student,
postdoc, and professor) are represented on Twitter. We extract this information
by searching for the following keywords in profile descriptions:
\begin{itemize}
\item student: \emph{phd student, phd candidate, graduate student,
grad student, doctoral student};
\item postdoc: \emph{postdoc, post-doc, postdoctoral};
\item professor: \emph{assistant professor, assistant prof, asst prof,
associate professor, associate prof, assoc prof, professor, prof, faculty}.
\end{itemize}
When more than one category are found, we choose the one that appears first. We
identify $3,705$ students, $1,030$ postdocs, and $5,326$ professors. This
indicates that many professors disclose their professional information on
Twitter.

\textbf{Community structure:} We understand the follower network from the
mesoscopic view---community structure. Analysis of communities helps us
understand how scientists' following activities are organized and what the
scholarly communities online are. These results will further advance our
understanding of the role of disciplines in the interactions between scientific
communities and of the comparisons with offline collaboration or citations
networks.

\begin{figure}[!h]
\includegraphics[trim=0mm 0mm 0mm 0mm, width=\columnwidth]{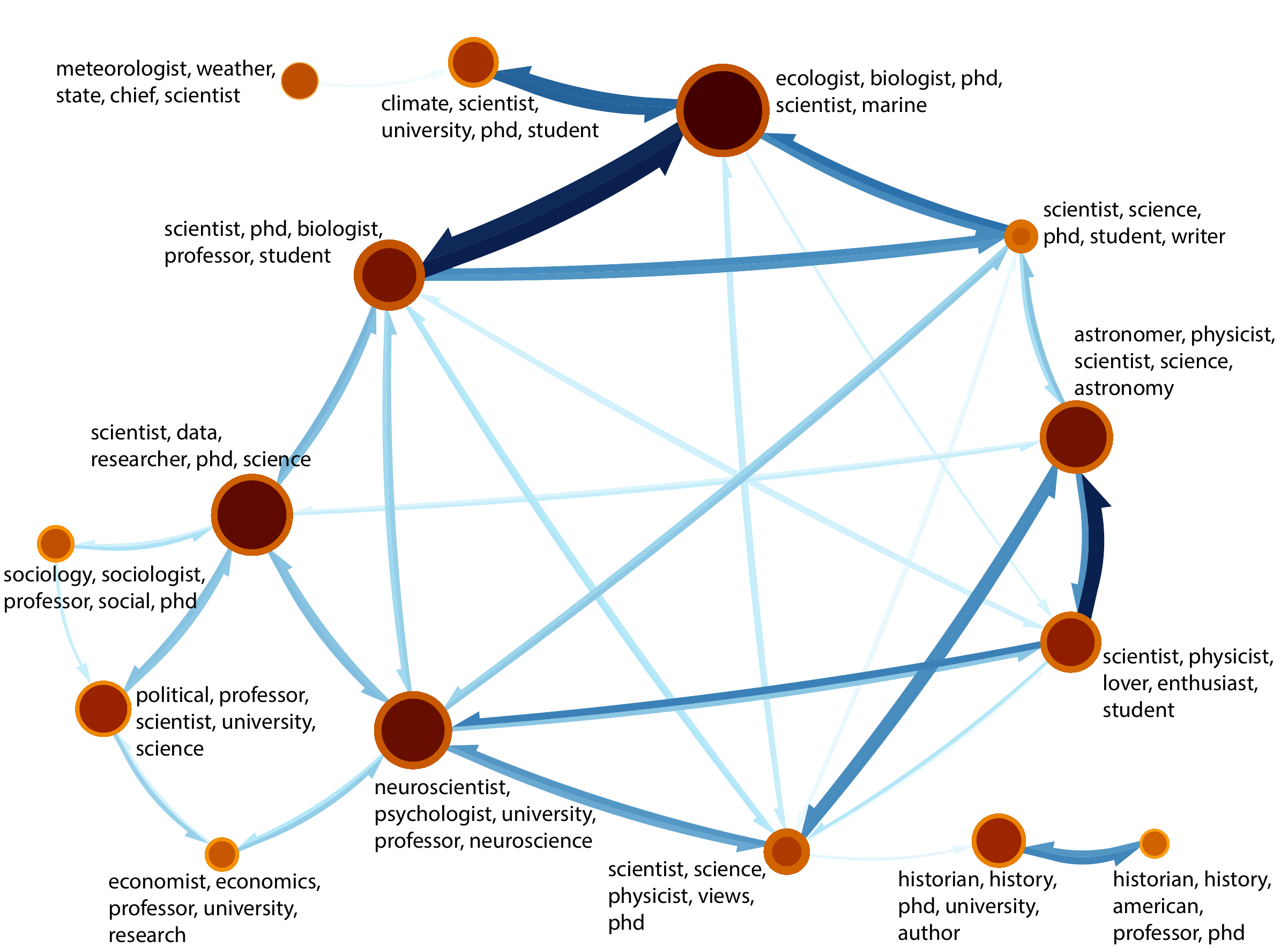}
\caption{
{\bf Network of communities.}}
\label{fig:follower-community}
\end{figure}

To identify communities in the follower network, we employed the Infomap
algorithm~\cite{Rosvall-infomap-2008} and identified $343$ communities with
more than $10$ nodes. Fig~\ref{fig:follower-community} shows the network
between the top $15$ communities. The number of links is set as the minimum
value that keeps the network connected. To understand what these communities
are, we count the appearance of individual words (excluding stop-words (a, and,
of, the, in, at, to, i, for, your, on, are, my, own, with)) in the profile
descriptions of users in each community. We use the top five most appeared
words to label each community, as showed in Fig~\ref{fig:follower-community}.
We can see that scientists seem to organize based on disciplines. They follow
other scientists in their own scientific communities. The two communities that
are composed with ecologists and biologists are tightly connected with each
other. This is also the case for (1) astronomers and physicists, and (2)
political scientists, economist, and sociologist. In
Table~\ref{tab:community-top-nodes}, we report the top scientists in each
community based on their PageRank.

\clearpage

\clearpage

\begin{table}[!ht]
\centering
\caption{
{\bf Scientist occupations from 2010 Standard Occupational Classification
released by US Department of Labor.}}
\begin{tabular}{c | r}
\hline
\multicolumn{2}{c}{15-1100 Computer Occupations}\\ \hline
15-1111 & Computer and Information Research Scientists \\ \hline
\multicolumn{2}{c}{15-2000 Mathematical Science Occupations}\\ \hline
15-2021 & Mathematicians \\
15-2041 & Statisticians \\
\hline \hline
\multicolumn{2}{c}{19-1000 Life Scientists}\\ \hline
19-1011 & Animal Scientists  \\
19-1012 & Food Scientists \\
19-1013 & Soil and Plant Scientists  \\
19-1021 & Biochemists and Biophysicists \\
19-1022 & Microbiologists \\
19-1023 & Zoologists and Wildlife Biologists \\
19-1031 & Conservation Scientists \\
19-1041 & Epidemiologists  \\ 
19-1042 & Medical Scientists \\ \hline
\multicolumn{2}{c}{19-2000 Physical Scientists}\\ \hline
19-2011 & Astronomers \\
19-2012 & Physicists  \\
19-2021 & Atmospheric and Space Scientists \\
19-2031 & Chemists \\
19-2032 & Materials Scientists \\
19-2041 & Environmental Scientists \\
19-2042 & Geoscientists \\
19-2043 & Hydrologists \\ \hline
\multicolumn{2}{c}{19-3000 Social Scientists and Related Workers}\\ \hline
19-3011 & Economists \\
19-3031 & Clinical, Counseling, and School Psychologists \\
19-3032 & Industrial-Organizational Psychologists \\
19-3041 & Sociologists \\
19-3091 & Anthropologists and Archeologists \\
19-3092 & Geographers \\
19-3093 & Historians \\
19-3094 & Political Scientists \\
\hline
\end{tabular}
\label{tab:sci-occu}
\end{table}

\begin{table}[!ht]
\centering
\caption{
{\bf Top scientist titles from profile descriptions.}}
\begin{tabular}{l r | l r}
\hline
Discipline & Users & Discipline & Users \\ \hline
Psychologist & 3379 & Sociologist & 538 \\
Historian & 2826 & Astronomer & 463 \\
Physicist & 2561 & Social scientist & 343 \\
Nutritionist & 2468 & Mathematician & 333 \\
Computer scientist & 1089 & Linguist & 320 \\
Archaeologist & 919 & Geographer & 319 \\
Political scientist & 891 & Epidemiologist & 294 \\
Biologist & 866 & Genealogist & 254 \\
Meteorologist & 818 & Geologist & 253 \\
Ecologist & 698 & Chemist & 242 \\
Neuroscientist & 665 & Astrophysicist & 236 \\
Economist & 661 & Microbiologist & 214 \\
Statistician & 599 & Environmental scientist & 208 \\
Clinical psychologist & 576 & Evolutionary biologist & 194 \\
Anthropologist & 546 & Pathologist & 177 \\
\hline
\end{tabular}
\label{tab:top-disci-profile}
\end{table}

\begin{table}[!ht]
\centering
\caption{
{\bf Top scientist titles from Twitter list names.}}
\begin{tabular}{l r | l r}
\hline
Discipline & Users & Discipline & Users \\ \hline
Psychologist & 4663 & Epidemiologist & 387 \\
Historian & 3371 & Geographer & 357 \\
Physicist & 2859 & Geologist & 344 \\
Nutritionist & 2510 & Evolutionary biologist & 336 \\
Archaeologist & 1183 & Genealogist & 298 \\
Sociologist & 996 & Social scientist & 281 \\
Economist & 955 & Ecologist & 193 \\
Biologist & 889 & Geoscientist & 188 \\
Meteorologist & 824 & Social psychologist & 181 \\
Astronomer & 768 & Pathologist & 172 \\
Political scientist & 756 & Astrophysicist & 159 \\
Anthropologist & 629 & Mathematician & 148 \\
Statistician & 594 & Microbiologist & 143 \\
Neuroscientist & 571 & Entomologist & 126 \\
Linguist & 476 & Chemist & 121 \\
\hline
\end{tabular}
\label{tab:top-disci-list}
\end{table}

\clearpage

\begin{table}[!ht]
\centering
\caption{
{\bf Top scientists in the follower, retweet, and mention networks between
scientists by in-degree $d_{\leftarrow}$ or in-strength $s_{\leftarrow}$,
PageRank ($PR$), and $k$-core number.}}
\begin{tabular}{c | l r | r r r | r r r}
\hline
$d_{\leftarrow}$ & \sname{neiltyson}, \sname{RichardDawkins}, \sname{sapinker}, \sname{phylogenomics}, \sname{donttrythis} \\
$PR$ & neiltyson, RichardDawkins, sapinker, \sname{SamHarrisOrg}, \sname{paulbloomatyale} \\
$k$  & \sname{randal\_olson}, \sname{Write4Research}, \sname{zacharyapte}, \sname{abcsoka}, \sname{ballenamar} \\\hline
$s_{\leftarrow}$ & neiltyson, \sname{AstroKatie}, \sname{elakdawalla}, phylogenomics, \sname{WhySharksMatter} \\
$PR$ & neiltyson, AstroKatie, \sname{conradhackett}, RichardDawkins, elakdawalla \\
$k$ & phylogenomics, \sname{surt\_lab}, \sname{duffy\_ma}, \sname{SciBry}, \sname{ethanwhite} \\ \hline
$s_{\leftarrow}$ & neiltyson, phylogenomics, RichardDawkins, WhySharksMatter, \sname{AtheneDonald} \\
$PR$ & neiltyson, RichardDawkins, sapinker, elakdawalla, phylogenomics \\
$k$ & \sname{raulpacheco}, \sname{CMBuddle}, \sname{mocost}, \sname{imascientist}, \sname{davenuss79} \\
\hline
\end{tabular}
\label{tab:centrality}
\end{table}

\begin{table}[!ht]
\centering
\caption{
{\bf Top users in each community.}}
\begin{tabular}{r | l l l}
\hline
   & Top users \\ \hline
1  & \sname{JacquelynGill}, \sname{GlobalEcoGuy}, \sname{SylviaEarle} \\
2  & \sname{zephoria}, \sname{EdwardTufte}, \sname{hmason} \\
3  & \sname{paulbloomatyale}, \sname{danariely}, \sname{deevybee} \\
4  & \sname{elakdawalla}, \sname{seanmcarroll}, \sname{AstroKatie} \\
5  & \sname{phylogenomics}, \sname{EricTopol}, \sname{JCVenter} \\
6  & \sname{neiltyson}, \sname{RichardDawkins}, \sname{sapinker} \\
7  & \sname{kinggary}, \sname{FukuyamaFrancis}, \sname{BrendanNyhan} \\
8  & \sname{holland\_tom}, \sname{JamesThorne2}, \sname{EZuelow} \\
9  & \sname{MichaelEMann}, \sname{KHayhoe}, \sname{ClimateOfGavin} \\
10 & \sname{jimalkhalili}, \sname{DrAliceRoberts}, \sname{RogerHighfield} \\
11 & \sname{JimCantore}, \sname{DrShepherd2013}, \sname{reedtimmerTVN} \\
12 & \sname{conradhackett}, \sname{alondra}, \sname{lisawade} \\
13 & \sname{TimHarford}, \sname{R\_Thaler}, \sname{CassSunstein} \\
14 & \sname{deborahblum}, \sname{kejames}, \sname{KateClancy} \\
15 & \sname{wcronon}, \sname{TomSugrue}, \sname{samueljredman} \\
\hline
\end{tabular}
\label{tab:community-top-nodes}
\end{table}

\end{document}